\documentclass{aa}
\usepackage{amsmath,amstext}
\usepackage[T1]{fontenc}
\usepackage{txfonts}
\usepackage[colorlinks=true,linkcolor = blue, citecolor=blue]{hyperref}
\usepackage[figure,figure*]{hypcap}

\usepackage{commath, mathrsfs}
\usepackage{color, upgreek}

\usepackage{mhchem}

\newcommand{\Rp}{R_{\rm p}}
\newcommand{\Rs}{R_{\rm s}}
\newcommand{\Mp}{M_{\rm p}}
\newcommand{\MJ}{{\,M_{\rm J}}}
\newcommand{\RJ}{{\,R_{\rm J}}}
\newcommand{\RB}{R_\mathrm{B}}
\newcommand{\Rhill}{R_\mathrm{Hill}}

\newcommand{\eV}{\,{\rm eV}}

\newcommand{\kms}{\,{\rm km\,s^{-1}}}

\newcommand{\AU}{ \, {\rm au}}
\newcommand{\au}{\AU}

\newcommand{\dd}{{\rm d}}

\newcommand{\cm}{{\, \rm cm}}

\newcommand{\K}{{\rm K}}

\newcommand{\braket}[1]{\left(#1\right)}

\newcommand{\e}[1]{\times 10^{#1}}
\newcommand{\Kelvin}{{\rm \, K}}

\begin{document}

\title{
Physically motivated analytic model of energy efficiency for EUV-driven atmospheric escape of close-in exoplanets
}
\author{H. Mitani
\inst{1,2} \thanks{E-mail: hiroto.mitani@uni-due.de}
\and
R. Nakatani
\inst{3}
\and
R. Kuiper
\inst{1}
}
\institute{
Faculty of Physics, University of Duisburg-Essen, Lotharstra{\ss}e 1, D-47057 Duisburg, Germany
\and
Department of Physics, School of Science, The University of Tokyo, 7-3-1 Hongo, Bunkyo, Tokyo 113-0033, Japan
\and
Dipartimento di Fisica, Universit\`a degli Studi di Milano, Via Celoria, 16, I-20133 Milano, Italy
}
\titlerunning{Analytic model of EUV-driven outflow}

\abstract
{Extreme Ultraviolet (EUV) driven atmospheric escape is a key process in the atmospheric evolution of close-in exoplanets. In many evolutionary models, the energy-limited mass-loss rate with a constant efficiency (typically $\sim10\%$) is assumed for calculating the mass-loss rate. However, hydrodynamic simulations have demonstrated that this efficiency depends on various stellar and planetary parameters. Comprehending the underlying physics of the efficiency is essential for understanding planetary atmospheric evolution and recent observations of the upper atmosphere of close-in exoplanets. We introduce relevant temperatures and timescales derived from physical principles to elucidate the mass-loss process. Our analytical mass-loss model is based on phenomenology and consistent across a range of planetary parameters. We compare our mass-loss efficiency and the radiation hydrodynamic simulations. The model can predict efficiency in both energy-limited and recombination-limited regimes.
We further apply our model to exoplanets observed with hydrogen absorption (Ly$\alpha$ and H$\alpha$). Our findings suggest that Ly$\alpha$ absorption is detectable in planets subjected to intermediate EUV flux; under these conditions, the escaping outflow is insufficient in low-EUV environments, while the photoionization timescale remains short in high-EUV ranges. Conversely, H$\alpha$ absorption is detectable under high EUV flux conditions, facilitated by the intense Ly$\alpha$ flux exciting hydrogen atoms. According to our model, the non-detection of neutral hydrogen can be explained by a low mass-loss rate and is not necessarily due to stellar wind confinement or the absence of a hydrogen-dominated atmosphere in many cases. This model assists in identifying future observational targets and explicates the unusual absorption detection/non-detection patterns observed in recent studies.}


\maketitle

\section{Introduction}
Extreme ultraviolet (EUV) radiation, with energy greater than approximately $13.6\mathrm{\, eV}$, from host stars plays a critical role in heating the upper atmospheres of close-in exoplanets. This occurs through the photoionization of hydrogen atoms. Such heating is a key driver of hydrodynamic escape \citep{Lammer_2003,Yelle_2004,Murray-Clay_2009, Owen_2012, Kubyshkina_2018b}, which significantly impacts the evolution of close-in exoplanets and influences various exoplanetary characteristics like the sub-Jovian desert and mass-radius gap \citep{Kurokawa_2014, Mazeh_2016, Fulton_2017, Owen_2017, Gupta_2019}. Numerous transit observations have detected escaping atmospheres around close-in exoplanets, typically through Ly$\alpha$ absorption due to its large transit depth \citep{Vidal-Madjar_2003,Lecavelier_2010,Kulow_2014,Ehrenreich_2015}. More recent studies have also used H$\alpha$ \citep[e.g.][]{Cauley_2017,Yan_2021} and the helium triplet \citep[e.g.][]{Oklopcic_2018, Spake_2018, Allart_2023})to observe upper atmospheres using ground-based telescopes. Despite the intense radiation expected to heat these upper atmospheres and drive hydrodynamic escape, some close-in planets show no helium absorption \citep{Bennett_2023, Allart_2023}. This absence could be explained by several factors: the planets might have already lost their hydrogen-dominated atmospheres \citep[e.g.][]{Zhang_2022}, strong stellar winds could be confining the atmospheres and reducing absorption signals \citep{Carolan_2020, Mitani_2022}, or the escaping outflow might be insufficient due to low EUV flux. Understanding the physics behind the atmospheric thermo-chemical structure is crucial for interpreting these observational trends.

There are several types of hydrodynamic escape driven by radiation from the host stars \citep{Murray-Clay_2009,Owen_2019}. If the radiation is not intense, the system becomes energy-limited, in which the mass-loss rate of the atmosphere is almost proportional to the flux of EUV radiation. Conversely, in the presence of intense ultraviolet radiation, radiative cooling regulates the gas temperature, and the system becomes recombination-limited. In this regime, the efficiency of mass loss decreases, and the mass-loss rate is not proportional to the flux but is nearly proportional to the square root of the flux. The energy efficiency varies with the escape regime and shapes the history of planetary mass loss. In many simulations of planetary evolution, the assumption of an energy-limited mass-loss rate with a constant efficiency is prevalent \citep[e.g.][]{Lopez_2013,Fujita_2022}. However, this simplification overlooks the variability of efficiency, which depends on several stellar and planetary parameters, thus not remaining constant. To improve the accuracy of planetary evolution models, it is necessary to understand the criteria that determine this efficiency. The physical conditions that govern these regimes are not well understood but depend on planetary mass, radius, surface temperature/density, orbital distance, stellar mass, stellar EUV luminosity, and EUV spectral hardness. Previous analytical models \citep{Kubyshkina_2018a} were based on a fit to simulations and did not derive these dependencies from a first-principles approach. The recent analytical model of hydrogen tails and observational signatures by \citet{Owen_2023} has shown the dependence of photoionization and mass-loss outflow on the observational absorption depth, treating the mass-loss rate as a free parameter. 

We develop an analytical model of the mass-loss rates in a first-principle approach to deepen our understanding of the physics of EUV-driven atmospheric escape depending on basic stellar and planetary parameters. We examine the physical conditions to analytically characterize the environment and establish simple criteria that can be used for planetary evolution models and for interpreting observations of atmospheric escape.

This paper is organized as follows.
In Section~\ref{sec:Physics}, we outline the physical timescales and temperatures that govern the physics of atmospheric escape driven by EUV photoionization heating.
In Section~\ref{sec:Analytic}, we present the analytic mass-loss rate and mass-loss efficiency applicable across a general parameter range, including energy-limited and recombination-limited regimes.
In Section~\ref{sec:classification}, we provide the comparison between our model and observed close-in planets. 
In Section~\ref{sec:discussion}, we give some further discussion of our model. In Section~\ref{sec:conclusion}, we present the conclusions.

\section{Essential Parameters}
\label{sec:Physics}
The dynamical and thermal state of the EUV-driven escaping gas is essentially governed by photoheating and gravity.
To investigate the basic physics of EUV-driven atmospheric escape, we introduce key quantities that describe these physical effects. 

We define the characteristic temperature $T_{\rm ch}$ \citep{1996_Woods, 1983_Begelman,Nakatani_2024} for a measure of the magnitude of photoheating as
\begin{equation}
k T_{\rm ch} = \frac{\Gamma_{\rm EUV}}{ c_p / \mu m_{\rm H}} \frac{\min(\Rhill,\RB) - R_{\rm EUV}}{c_{\rm ch}}, 
\label{eq:cch}
\end{equation}
where $\Gamma_{\rm EUV}$ represents the specific EUV photoheating rate, $\Rhill$ is the Hill radius, $R_{\rm EUV}$ is the effective planetary radius where the optical depth to EUV photons is unity, $\RB=G\Mp/(2c_{\rm s}^2)$ is the Bondi radius where $c_{\rm s}$ is the isothermal sound speed of the atmosphere, $k$ is the Boltzmann constant, $c_p = \gamma/(\gamma-1)$ is the normalized specific heat at constant pressure with the specific heat ratio $\gamma$, $\mu$ is the mean molecular weight, $c_{\rm ch}$ is the sound speed of the gas at the characteristic temperature, and $m_{\rm H}$ is the hydrogen atomic mass. 
Note that the sound speed $c_{\rm ch}$ on the right-hand-side (RHS) depends on the characteristic temperature $T_{\rm ch}$.  
The RHS represents the total deposited photo-energy within the typical sound crossing time. We use $\min(\Rhill,\Rs) - R_{\rm EUV}$ as the typical length scale of the streamline where the bulk of heating occurs, as the photoheating outside the sonic point does not affect the thermal structure within the sonic point and if the Hill radius is smaller than the Bondi radius, the stellar gravity boosts the gas velocity to sonic speed around the Hill radius.
Hence, the characteristic sound speed is defined as:
\[
c_{\rm ch} = \left(\frac{\Gamma_{\rm EUV} \Rp'}{c_p}\right)^{1/3}. 
\]
where  $\Rp' = \min(\Rhill,\RB) - R_{\rm EUV}$.
The characteristic temperature indicates how rapid the photoheating is compared to the dynamical timescale but does not necessarily give the typical temperature of the photoheated gas. For instance, under conditions of high EUV irradiation, the characteristic temperature may exceed a million Kelvin. However, the actual atmospheric temperature is kept lower due to radiative cooling processes, which effectively limit the gas temperature.
In such a planet with high characteristic temperature, the fraction of the neutral hydrogen is not unity and the characteristic temperature should be lowered. However, the gas temperature reaches the equilibrium temperature as demonstrated in Section~\ref{sec:mass-loss} and the characteristic temperature as in Eq.~\ref{eq:cch} is practically useful to understand the gas temperature. 

The heating timescale $t_{\rm h}$ is defined as 
\[
    t_{\rm h} = \frac{h}{\Gamma_{\rm EUV}} = \frac{\Rp'}{c_{\rm ch}}, 
\]
where $h$ is the specific enthalpy.

The specific heating rate $\Gamma_{\rm EUV}$ is 
\begin{equation}
\begin{split}
    \Gamma_{\rm EUV} & =  \frac{y_{\rm HI} }{m} \int \sigma_\nu (h\nu - h\nu_1) F_{\nu, 0} \exp(-\sigma_\nu N_{\rm HI}) \, \dd \nu \\
    &   =   \frac{y_{\rm HI}  }{m} F_0 \delta 
            \langle \sigma \rangle
            \langle \Delta E \rangle_i  
\end{split}
\label{eq:gamma}
\end{equation}
where $m$ denotes the gas mass per hydrogen nucleus, $\sigma_\nu$ is the absorption cross-section of neutral hydrogen for EUV frequency $\nu$, $y_{\rm HI}$ is the atomic hydrogen abundance, and
$\delta$ is the ratio of the attenuated flux to the unattenuated flux ($F_{\nu, 0}$) at a given \ion{H}{I} column density ($N_{\rm HI}$), 
\[
\delta (N_{\rm  HI})\equiv \dfrac{\int \dd \nu \,  F_{\nu,0} \exp \braket{- \sigma_\nu N_{\rm HI}}}{\int \dd \nu \,   F_{\nu,0} }. 
\]
$\langle \sigma \rangle $ is the average cross section of the ionizing photons reaching $N_{\rm HI}$
\[
\langle \sigma \rangle (N_{\rm  HI}) \equiv \frac{\int \dd \nu \,  \sigma_\nu F_{\nu,0} \exp \braket{- \sigma_\nu N_{\rm HI}}}{\int \dd \nu \,   F_{\nu,0} \exp \braket{- \sigma_\nu N_{\rm HI}}} ,
\]
and $\langle \Delta E \rangle_i$ is the average deposited energy per ionization at $N_{\rm HI}$
\[
\langle \Delta E \rangle_i (N_{\rm  HI}) \equiv \frac{\int \dd \nu\,  \sigma_\nu (h\nu - h\nu_1) \frac{\Phi_\nu \exp(-\sigma_\nu N_{\rm HI})}{4\pi r^2}}{\int \dd \nu \,  \sigma_\nu F_{\nu,0} \exp \braket{- \sigma_\nu N_{\rm HI}}}
\]
$\Phi_{\nu}$ is the specific stellar EUV emission rate and $r$ is the distance to the planet from the host star.
All of $\delta$, $\langle \sigma \rangle$, and $\langle \Delta E \rangle_i$ are dependent only on $N_{\rm HI}$ and a priori calculable, given the EUV spectrum. 
The former two are monotonically decreasing functions, while the other is a monotonically increasing function. 
The product $ \delta \langle \sigma\rangle \langle E\rangle_i$ is a monotonically decreasing function with respect to $N_{\rm HI}$, which is also evident by the $N_{\rm HI}$-derivative of $\Gamma_{\rm EUV}$.
For later convenience, we define $\sigma_0 \equiv \left.\langle \sigma \rangle\right|_{N_{\rm HI} = 0}$ and $\Delta E_0 \equiv \left. \langle \Delta E \rangle_i\right|_{N_{\rm HI} = 0}$. 

We can rewrite $\Gamma_{\rm EUV}$ in Eq.~\eqref{eq:gamma} as:
\begin{equation}
    \Gamma_{\rm EUV} = \frac{y_{\rm HI}}{m}F_0\sigma_0\Delta E_0\chi_e
    \label{eq:gamma_euv}
\end{equation}
where
\[
    \chi_e = \delta\frac{\langle \sigma\rangle}{\sigma_0}\frac{\langle \Delta E \rangle_i}{\Delta E_0}
\]
$\chi_e$ is a function of column density and the range is 0 to 1. In this study, we neglect the attenuation because the atmospheric density structure is highly dependent on the radius. For realistic spectra, this assumption is not necessary but constructing a model based on the simplification is a step towards a full understanding of atmospheric escape driven by photoheating and practically useful to understand the mass-loss rate of close-in exoplanets.

Correspondingly, the characteristic sound speed is 
\[
\begin{split}
c_{\rm ch}  &   =       
            6.24\kms 
            \braket{\frac{y_{\rm HI}}{0.5}}^{1/3}
            \braket{\frac{m}{1.4m_{\rm H}}}^{-1/3}  \\  
    &   \times            
            \braket{\frac{\Phi}{10^{41}\sec^{-1}}}^{1/3}
            \braket{\frac{r}{1\au}}^{-2/3}
            \braket{\frac{\Rp'}{10^{10}\cm}}^{1/3}\\
    &   \times            
            \braket{\frac{c_p}{5/2}}^{-1/3}
            \braket{\frac{\sigma_0 }{5\e{-18}\cm^2}}^{1/3} 
            \braket{\frac{ \Delta E_0}{1\eV}}^{1/3} 
            \chi_e.
\end{split}
\]
where $\Phi$ is EUV luminosity of the host star.

From the balance of photoheating and cooling, the gas has an equilibrium temperature $T_{\rm eq}$, which is typically $\approx10^4 \Kelvin$ for EUV-heated gas, with the corresponding sound speed $c_{\rm eq} \approx 10\kms$. Note that the equilibrium temperature is not the planetary surface temperature, and that the atmospheric temperature cannot significantly exceed the equilibrium temperature. The equilibrium temperature may depend on the metal abundance of the atmosphere, as metal species serve as coolants. We discuss the effects of metallicity in Section~\ref{sec:metal}.

The planetary gravity also plays an important role in the atmospheric escape process.
The gravitational timescale is accordingly
\[
    t_{\rm g} = \sqrt{\frac{\Rp^3}{G\Mp}}
\]
We also define the gravitational temperature
\begin{equation}
    T_{\rm g} = \frac{G\Mp \mu m_{\rm H}}{c_p\Rp k},
    \label{eq:T_g}
\end{equation}
which characterizes the strength of planetary gravity and is compared with $T_{\rm eq}$ and $T_{\rm ch}$. Essentially, the gas with temperatures hotter than $T_\mathrm{g}$ does not feel planetary gravity, and fast winds can be driven. If the characteristic temperature is lower than the gravitational temperature, the gas is significantly inhibited by gravity, and the winds tend to be subsonic ($v_{\rm gas} < c_{\rm g}$ where $v_{\rm gas}$ is the gas velocity and $c_{\rm g}$ is the sound speed at the gravitational temperature) with temperatures of the order of the gravitational temperature, as discussed in \cite{Nakatani_2024}. 

The neutral hydrogen fraction also plays a crucial role in interpreting the observed hydrogen absorption. 
The balance between photoionization and recombination determines the neutral hydrogen fraction because the dynamical timescale is usually longer than the photoionization timescale in the case of close-in exoplanets. We can define the photoionization/recombination timescales as:
\begin{eqnarray}
t_{\rm ion} &=& \frac{1}{F_0\sigma_0}, \label{eq:t_ion}\\
t_{\rm rec} &=& \frac{1}{n_{\rm H} \alpha_{\rm rec}},
\label{eq:t_rec}
\end{eqnarray}
where $t_{\rm ion}$ is the photoionization timescale, $t_{\rm rec}$ denotes the recombination timescale, and $n_{\rm H}$ is the number density of hydrogen atoms. $\alpha_{\rm rec} \sim 2.7\times10^{-13}\mathrm{\, cm^3/s}$ is the radiative recombination rate coefficient for hydrogen ions. 
The recombination coefficient is temperature-dependent; however, in many cases, we find that this dependence has a minimal impact on our results. This is primarily because the recombination coefficient scales approximately with the square root of temperature, meaning that temperature variations result in only minor adjustments to the ratio of the relevant timescales. It is important to note that the photoionization timescale is influenced by the EUV flux (which is a function of both EUV luminosity and orbital distance), while the recombination timescale is determined by the gas density. For planetary outflows, these calculations can be based on the physical conditions at the base of the flow.

Up to this point, we have introduced $t_{\rm h}$, $t_{\rm ion}$, $t_{\rm g}$, and $t_{\rm rec}$ as relevant timescales, $T_{\rm ch}$, $T_{\rm eq}$, and $T_{\rm g}$ as relevant temperatures,and $c_{\rm ch}$, $c_{\rm g}$, and $c_{\rm eq}$ as relevant sound speeds.

If $T_{\rm ch} > T_{\rm eq}$, photoheating is so rapid that it can increase the gas temperature $T_{\rm gas}$ to $T_{\rm eq}$ before flowing out to a distance of $\sim R_\mathrm{hill}$. In this regime,  
$T_{\rm gas}\approx T_\mathrm{eq}$.
The ratio between $T_{\rm ch}$ and $T_{\rm eq}$ is 
\[
    \frac{T_{\rm ch}}{T_{\rm eq}}
    =   \frac{c_{\rm ch}^2}{c_{\rm eq}^2}
    =   \braket{\frac{\Gamma_{\rm EUV} \Rp'}{c_{\rm eq}^3 c_p}} ^{2/3}
    =    \braket{\frac{\Gamma_{\rm EUV} \Rp' G \Mp }{R_{\rm g} c_p c_{\rm eq}^5 }} ^{2/3}. 
\]
where $R_\mathrm{g}$ is the gravitational radius defined as 
\[
    R_{\rm g} = \frac{G\Mp}{c_{\rm eq}^2}.
\]
The above ratio is written to 
\[
    \frac{T_{\rm ch}}{T_{\rm eq}} =  \xi^{2/3} \braket{\frac{ \Gamma_{\rm EUV} G \Mp }{c_p c_{\rm eq}^5}} ^{2/3},
\]
where $\xi \equiv \Rp'/ R_{\rm g}$. 
The second factor in the RHS defines the critical flux
\[
\frac{\Gamma_{\rm EUV} G \Mp}{c_p c_{\rm eq}^5}
=\frac{F_0}{F_{\rm cr}} \chi_e y_{\rm HI}
\]
where
\begin{equation}
\begin{split}
    F_{\rm cr} & \equiv \frac{ mc_p c_{\rm eq}^5} {G\Mp \sigma_0 \Delta E_0}\\
    &   \approx
            5.8\e{12}\cm^{-2} \sec^{-1}
            \braket{\frac{m}{1.4m_{\rm H}}}  
            \braket{\frac{c_{\rm eq}}{10\kms}} ^5
    \\      &   \times            
            \braket{\frac{\Mp}{1\MJ}}^{-1}
            \braket{\frac{\sigma_0}{5\e{-18}\cm^2}}^{-1}
            \braket{\frac{ \Delta E_0}{1\eV}}^{-1}
            \braket{\frac{c_p}{5/2}},
\end{split}
\label{eq:critical_flux}
\end{equation}
Note that the $T_{\rm ch}/T_{\rm eq}$ ratio determines the physical regime, while the critical EUV flux does not. 

Similarly, the ratio of $T_\mathrm{ch}$ to $T_\mathrm{g}$ is calculated as
\[ 
    \frac{T_\mathrm{ch}}{T_\mathrm{g}}
    = \frac{c_\mathrm{ch}^2}{GM_\mathrm{p}/R_\mathrm{p}} 
    = \frac{T_\mathrm{ch}}{T_\mathrm{eq}}\frac{R_\mathrm{p}}{R_\mathrm{g}}
    = \braket{\frac{F_0}{F_\mathrm{cr}}\chi_e y_\mathrm{HI}}^{2/3}
    \xi^{2/3}\braket{\frac{R_\mathrm{p}}{R_\mathrm{g}}}. 
\]
Thus, the ratio of $R_\mathrm{p}$ to $R_\mathrm{g}$ determines whether the subspace where $T_\mathrm{ch} > T_\mathrm{eq}$ also satisfies $T_\mathrm{ch} > T_\mathrm{g}$, or vice versa. For relatively massive planets ($R_\mathrm{g} > R_\mathrm{p}$), the parameter space where $T_\mathrm{ch} > T_\mathrm{g}$ always satisfies $T_\mathrm{ch} > T_\mathrm{eq}$, implying that winds can be gravitationally inhibited even with temperatures of $T_\mathrm{eq}$.

\section{Analytic model}
\label{sec:Analytic}
Mass-loss rates determine the planetary evolution and the absorption signals.
Understanding the basic physics of the mass-loss driven by EUV-photoionization is of use for, e.g., finding the origins of peculiar observational signals.
In this section, we summarize our analytic model of mass-loss rates, which is built upon the essential parameters in Section~\ref{sec:Physics} and physical conditions that govern the mass-loss efficiency.
Our model is capable of seamlessly bridging between the energy-limited and recombination-limited mass-loss regimes. 
We provide the analytic formulae in Section~\ref{sec:mass-loss}. Additionally, we compare our model's predictions with results from radiation hydrodynamic simulations in Section~\ref{sec:simulation}. This comparison helps validate our analytical approach and provides insights into the dynamics of atmospheric escape under various stellar and planetary conditions.

\subsection{Mass-loss rates and efficiency}
\label{sec:mass-loss}
Mass loss driven by EUV photoionization heating is a crucial process in the evolution of close-in planets. Many planetary evolution studies rely on a simplified model where the mass-loss rate is calculated using an energy-limited formula \citep{1981_Watson, Erkaev_2007}:
\begin{equation}
    \dot{M} = \eta \frac{F_0\Rp^3}{GK\Mp}
    \label{eq:eta}
\end{equation}
Here $\dot{M}$ represents the mass-loss rate, $\eta$ is the efficiency of mass-loss, and $K$ is the factor accounting for stellar gravity \citep{Erkaev_2007}. While this formula provides a useful estimate, it usually assumes a constant efficiency ($\eta$), which simplifies the complex interplay of factors that can influence mass loss. In reality, $\eta$ can vary significantly based on the planetary and stellar characteristics. In this section, we formulate an analytic expression for the efficiency that depends on planetary and stellar parameters, using the relevant quantities introduced in Section~\ref{sec:Physics}.

In contrast to the energy-limited regime, radiative cooling plays an important role in the recombination-limited regime. In this regime, previous studies \citep{Murray-Clay_2009, Owen_2017} estimate the mass-loss rate by approximating the density profile within the sonic point to that in hydrostatic equilibrium:
\begin{equation}
    \dot{M} \sim \pi\Rs^2 c_{\rm s} m_{\rm H} n_{\rm base} \exp\left(2-\frac{2\Rs}{\Rp}\right),
    \label{eq:mdot}
\end{equation}
where $\Rs = \RB$, and fully ionized atmosphere ($\mu = 0.5$) with the equilibrium temperature ($T_{\mathrm{gas}} = T_\mathrm{eq}$,) are assumed to calculate the sound speed $c_{\rm s}$; $\pi\Rs^2$ has been applied instead of $4\pi\Rs^2$ to account for the day-side illumination effect.
The base density is given as $n_{\rm base}\sim(F_0/\alpha_{\rm rec}H)^{1/2}$ with $H$ being the pressure scale height, which is defined as in \cite{Owen_2016}:
\begin{equation}
    H = \min\left[\frac{\Rp}{3},\frac{c_{\rm s}^2 \Rp^2}{2G \Mp}\right]
\end{equation}
The base density formula, although originally derived under the assumption of a thin ionization front, primarily depends on the local recombination-ionization balance and the gas temperature rather than on the actual thickness of the front. Even with a thick ionization front, the gas near the base still reaches equilibrium between ionization and recombination if the recombination timescale is shorter than the flow timescale. 
The scale height depends on the gas temperature and we can estimate the base density not only for an atmosphere of $10^4\mathrm{\, K}$, but also for a relatively low-temperature atmosphere. If the EUV flux is weak on a weak gravity planet, such as a sub-Neptune, this assumption is incorrect and the mass-loss rate gets overestimated.

In this model, the effect of stellar gravity is neglected. To ensure consistency across different mass-loss models, we similarly ignore stellar gravity when calculating the efficiency of mass loss. Specifically, we adopt the approximation $K=1$ in our calculations to align with the simplifications inherent in the recombination-limited model. This approximation effectively assumes stellar gravity to be negligible. However, including the gravitational factor $K$ in the planetary mass or radius does not significantly alter the results.

Essentially, our model updates Eq.~\eqref{eq:mdot} by incorporating the variations in the gas temperature $T_\mathrm{gas}$, the ionization degree of gas, and the sonic point location $\Rs$, which depend on the stellar and planetary parameters. We use a single representative temperature, neglecting the spatial temperature profile.
We express the representative atmospheric temperature as follows:
\[
T_{\mathrm{gas}} = \min(T_{\rm eq}, \max(T_{\rm ch}, T_{\rm g}))
\]
This expression is derived from the following phenomenological considerations: the gas temperature does not exceed the equilibrium temperature due to the regulation by radiative cooling (Regime~A-1) otherwise, the gas temperature $T_{\mathrm{gas}}$ is basically determined by a characteristic temperature $T_{\rm ch}$ (Regime~A-2). On the other hand, if the photoheating is weak to yield $T_\mathrm{ch}$ lower than the gravitational temperature $T_{\rm g}$ (Regime~A-3), the deposited energy is charged to heat the gas to $T_\mathrm{g}$ at most. 
We note that strictly, $T_\mathrm{g}$ should be evaluated using $R_\mathrm{EUV}$ instead of $\Rp$ in Eq.~\eqref{eq:T_g}. However, since $\Rp \approx R_\mathrm{EUV}$ in this regime, using Eq.~\eqref{eq:T_g} does not affect the results.

The revisions to $T_\mathrm{gas}$ also update the sound speed ($c_\mathrm{s}$) accordingly.
\begin{equation}
    c_{\rm s}=\min(c_{\rm eq},\max(c_{\rm ch}, c_{\rm g})),
    \label{eq:corrected_cs}
\end{equation}
where $c_{\rm eq}$, $c_{\rm ch}$, and $c_{\rm g}$ are the sound speeds at equilibrium temperature, characteristic temperature, and gravitational temperature, respectively (cf. Table~\ref{tab:model1}). 
Note that $c_\mathrm{s}$ depends on the mean molecular weight ($\mu$) as well, which is determined once the ionization degree or, equivalently, the atomic hydrogen abundance is set. We approximate the atomic hydrogen abundance to $\min(t_\mathrm{ion}/t_\mathrm{rec}, 1)$, which in turn depends on $c_\mathrm{s}$ through the scale height $H$ incorporated in the base density.
Hence, $c_\mathrm{s}$ and $\mu$ must be determined simultaneously. 
The updates to $T_\mathrm{gas}$ and accordingly $c_\mathrm{s}$ are summarized in Table~\ref{tab:model1} along with the corresponding parameter spaces.
\begin{table*}[h]
\caption{Summary of our analytical model for sound speed regimes.}
\centering
\begin{tabular}{ccccc}
    \hline
    Regime & Description & Condition & Sound speed\\
    \noalign{\smallskip}
    \hline
    \noalign{\smallskip}
     A-1 & Radiative cooling thermostat &$T_{\rm ch} > T_{\rm eq}$ & $c_{\rm eq}$\\ 
     A-2 & $PdV$ work cooling & $T_{\rm eq} > T_{\rm ch}$ & $c_{\rm ch}$\\
     A-3 & Inhibited by planetary gravity  &$T_{\rm eq} > T_{\rm g} > T_{\rm ch}$ & $c_{\rm g}$\\
     \noalign{\smallskip}
     \hline
\end{tabular}
\label{tab:model1}
\end{table*}

As for the update to $\Rs$, it does not necessarily coincide with the Bondi radius $\RB$ in general, as implicitly assumed in the classical model (Eq.~\eqref{eq:mdot}). 
In the parameter space where the Bondi radius exceeds the Hill radius $\RB > \Rhill$, such as when the gas temperature is low or the stellar gravity is relatively strong, the stellar gravity becomes the primary accelerating source, and thus $\Rs\approx\Rhill$. 
In contrast, for the case where the Bondi radius approaches the wind-launching radius $\RB \rightarrow R_\mathrm{EUV}$, meaning a high gas temperature or weak planetary gravity, the winds are instantly accelerated to be supersonic after being launched at the base $R_\mathrm{EUV}$, resulting in $\Rs\approx R_\mathrm{EUV}$. 
Note that we are not interested in the case where EUV photons fail to penetrate within the Hill radius, $R_\mathrm{EUV} > \Rhill$, and that $\RB$ is constrained to not fall below $R_\mathrm{EUV}$ by definition (cf. Eq.\eqref{eq:cch}). 
We summarize the sonic point locations in Table~\ref{tab:model2}.
\begin{table*}[h]
\caption{Summary of our analytical model of radius conditions. The mass-loss rate can be given as Eq.~\eqref{eq:mdot1} or Eq.~\eqref{eq:mdot2}.}
\centering
\begin{tabular}{ccccc}
    \hline
    Regime & Sonic point $\Rs$ & Condition & Mass-loss rate\\
    \noalign{\smallskip}
    \hline
    \noalign{\smallskip}
     Classical picture & $\RB$ & $R_{\rm EUV}<\RB <R_{\rm hill}$  &$\dot{M}=\dot{M}_1(\RB)$\\
    Stellar gravity dominant &$\Rhill$& $R_\mathrm{EUV} <  R_{\rm hill} < \RB$ & $\dot{M}=\dot{M}_1(R_{\rm hill})$\\
     Instant acceleration at the base & $R_\mathrm{EUV}$ &$\RB \sim R_{\rm EUV}$  &$\dot{M}=\dot{M}_2(R_{\rm EUV})$\\
     \noalign{\smallskip}
     \hline
\end{tabular}
\label{tab:model2}
\end{table*}

The update to the sonic point locations results in two distinct expressions for $\dot{M}$, depending on whether the sonic point $\Rs$ lies at the base $R_\mathrm{EUV}$. 
If $R_\mathrm{EUV} < \RB$, the expression for $\dot{M}$ remains essentially the same as Eq.~\eqref{eq:mdot}:
\begin{equation}
    \dot{M}_1(\Rs) = \pi\Rs^2 v_{\rm gas} m_{\rm H} n_{\rm base} \exp\left[\frac{2\RB}{R_{\rm EUV}}\left(\frac{R_{\rm EUV}}{\Rs}-1\right)\right].
    \label{eq:mdot1}
\end{equation}
Here, we have replaced $\Rp$ in the original equation Eq.~\eqref{eq:mdot} with $R_{\rm EUV}$. 
On the contrary, for $\RB \rightarrow R_\mathrm{EUV}$, $R_{\rm EUV}$ now serves as the sonic point, and $\dot{M}$ is updated to
\begin{equation}
    \dot{M}_2(R_{\rm EUV}) = \pi R_{\rm EUV}^2 m_{\rm H} n_{\rm base} v_{\rm gas}  
    \label{eq:mdot2}
\end{equation}
This equation is the asymptotic form of Eq.\eqref{eq:mdot1} when $\RB \rightarrow R_\mathrm{EUV}$.

In Regimes~A-1 and A-2, the gas velocity $v_\mathrm{g}$ in Eqs.~\eqref{eq:mdot1} and \eqref{eq:mdot2} is set to $c_\mathrm{s}$ (Eq.~\eqref{eq:corrected_cs}) as in the original equation (Eq.~\eqref{eq:mdot}) . However, in Regime~A-3 (gravity-inhibited regime), we apply an empirical correction factor based on the simulation results of our 1D simulations and ATES-code to get more realistic mass-loss efficiencies for low EUV environments: $v_{\rm gas} = c_{\rm g} ({t_{\rm g}}/{t_{\rm h}})$. This adjustment is necessary because using a single representative temperature $T_\mathrm{g}$, while neglecting the spatial profile ($1/r$), is not valid in this regime, and the actual sound speed at the sonic point is significantly lower than the evaluated sound speed $c_{\rm g}$.

Finally, the mass-loss rates of Eqs.~\eqref{eq:mdot1} and \eqref{eq:mdot2} are fully determined once the effective planetary radius $R_\mathrm{EUV}$ is specified.

We set this radius as the radius where the density of the isothermal atmosphere with the surface temperature becomes the base density:

\begin{equation}
R_{\rm EUV} = \Rp \left(1+\frac{\Rp}{2 R'_{\rm B}} \log(\rho_{\rm base}/\rho_{\rm surf})\right)^{-1}
\label{eq:R_euv}
\end{equation}
where $R'_{\rm B}$ is the Bondi radius based on the planetary surface temperature, and $\rho_{\rm surf}$ is the density at the planet's surface. 

Note that we neglect the spectral dependence of $R_{\rm EUV}$ but the density structure of the inner atmosphere is strongly dependent on radius. The cross-section dependence of $R_{\rm EUV}$ has a minor effect on the mass-loss rate.

\begin{figure*}[h]
    \centering
    \includegraphics[width=19cm]{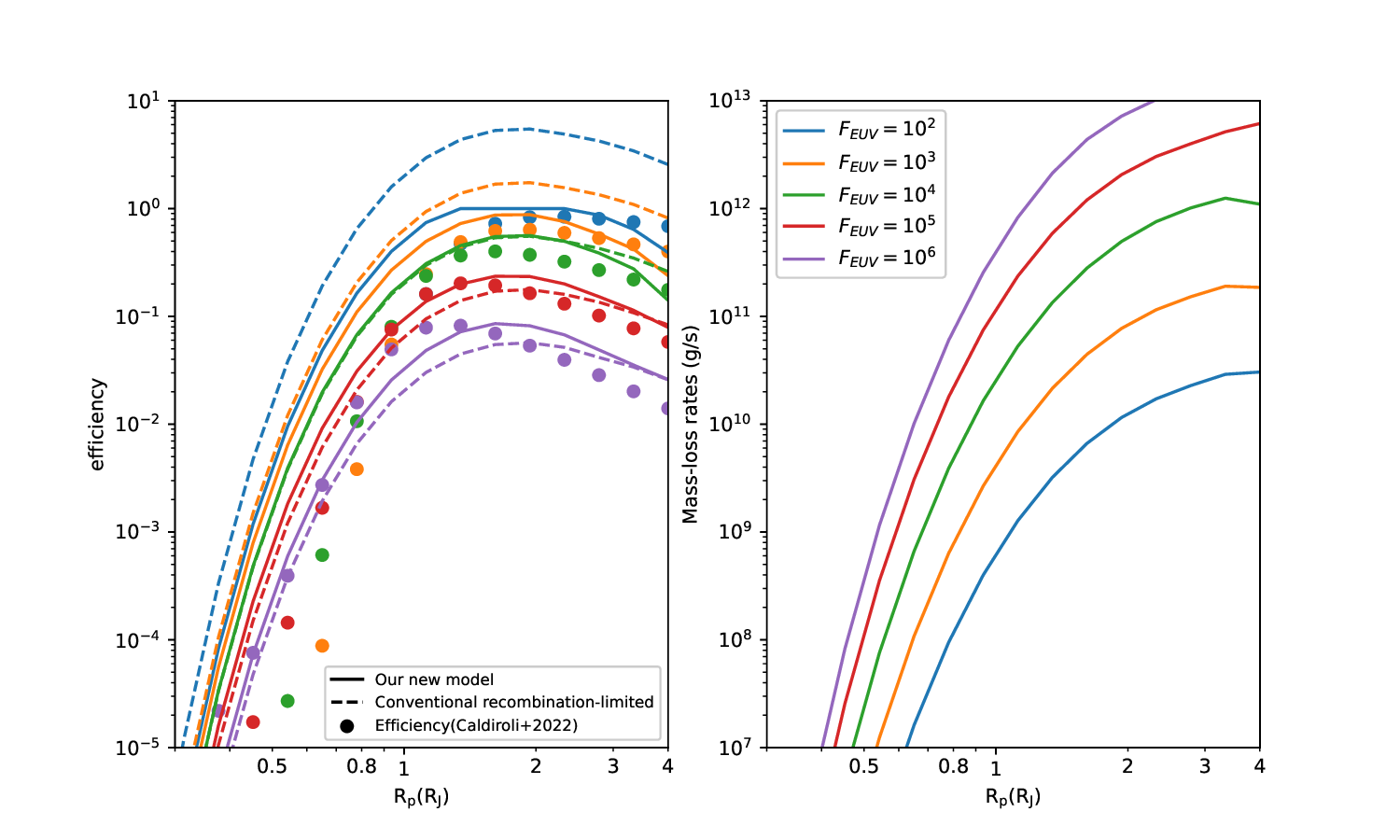}
    \caption{Estimated efficiencies (left panel) and mass-loss rates (right panel) for fixed planetary mass $\Mp=0.7\,\MJ$ but different EUV fluxes from $F_{\rm EUV} = 100\mathrm{\,erg/s/cm^2}$ (red) to $F_{\rm EUV} = 10^6\mathrm{\,erg/s/cm^2}$ (orange). The efficiency of the traditional recombination-limited approach is shown in dashed curves. In the intense EUV flux case, our model predictions are almost the same as the traditional approach (orange and purple). The fitted efficiencies from 1D detailed simulations by \citet{Caldiroli_2021} are shown as points. Some points of the strong-gravity case ($\Mp>20\,M_{\rm J}$) can not be seen in this panel due to the low efficiency.}
    \label{fig:massloss_m}
\end{figure*}

By applying Eqs.~\eqref{eq:eta}, \eqref{eq:mdot1} and \eqref{eq:mdot2}, we can semi-analytically calculate the mass-loss efficiency $\eta$ across any given parameter space (see also Table~\ref{tab:model2}).
In Figure~\ref{fig:massloss_m}, we present the calculated efficiencies for a fixed planetary mass. The conventional recombination-limited mass-loss rates sharply drop in regions of strong gravity (corresponding to smaller $\Rp$, or equivalently smaller $R_\mathrm{EUV}$ in terms of Eq.~\eqref{eq:R_euv}). This sharp decline is primarily due to a larger ratio of the sonic radius to the planetary radius, which causes the density at $\Rs$ to diminish exponentially (cf. Eq.~\eqref{eq:mdot1}). The efficiency of the conventional recombination-limited model tends to overestimate the efficiency in environments characterized by weak gravity. In cases of a thick ionization front, the base density can be overestimated, which poses significant challenges for low-gravity and low-EUV flux planets.
However, in such scenarios, the density profile becomes less steep with respect to the radius, thereby mitigating the impact of this overestimation.
Our model tends to overestimate the mass-loss rate in these cases. However, the estimated sound speed compensates for this overestimation, unlike the conventional recombination-limited mass-loss rate.

We also compare our efficiency with that derived from 1D simulations \citep{Caldiroli_2021}. We find that our efficiency more closely aligns with the simulation results in highly irradiated ($F_{\rm EUV}>10^4\mathrm{\,erg/s/cm^2}$) and relatively weak gravity ($\phi_p=G\Mp/\Rp < 2\times10^{13} \mathrm{\,erg/g}$) cases, compared to the conventional recombination-limited cases.

The slightly decreasing trend in the efficiencies for larger-sized planets is because $\eta$ is proportional to $\dot{M}\Rp^{-3}$ as in Eq.~\eqref{eq:eta}, with the mass-loss rate weakly depends on the radius (proportional to the square root of the radius as in Eq.~\eqref{eq:mdot2} and $\propto (1-2\Rs/\Rp)$ in Eq.~\eqref{eq:mdot} for larger $\Rp$).

\begin{figure*}[h]
    \centering
    \includegraphics[width=19cm]{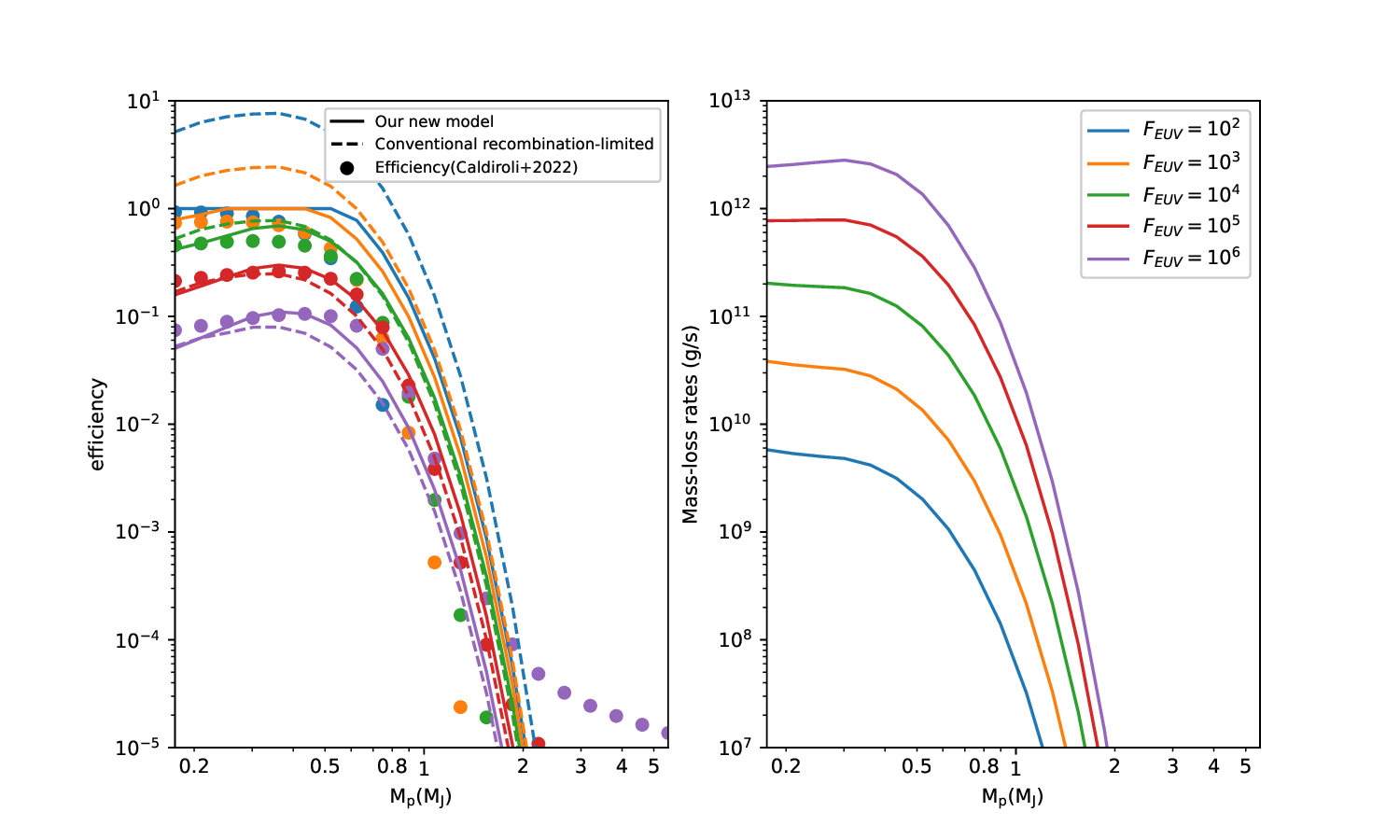}
    \caption{Same as Figure~\ref{fig:massloss_m} but here for a varying planetary mass (horizontal axis) and a fixed planetary radius $\Rp=\RJ$.}
    \label{fig:massloss_r}
\end{figure*}

Next, we discuss the dependence of mass-loss efficiency on planetary mass.
Figure~\ref{fig:massloss_r} presents the planetary mass dependence of the efficiencies, similar to Figure~\ref{fig:massloss_m} but for a fixed planetary radius. The global trend is also akin to that can be found in Figure~\ref{fig:massloss_m}; however, the dependence on gravity differs in high-irradiation and lower-gravity environments (left side of the curves), where the efficiency appears almost independent of planetary gravity. This is attributed to the gravity-dependent nature of the effective planetary radius, $R_{\rm EUV}$ (Eq.~\eqref{eq:R_euv}). Efficiency $\eta \propto \dot{M}\Mp$, and both $R_{\rm EUV}$ and the mass-loss rate increase for lower-mass planets. Near $\Mp \sim M_{\rm J}$, the Bondi radius approaches the Hill radius. We also find that the Hill radius affects the efficiency slightly ($<10\%$) for high-mass planets ($\Mp > M_{\rm J}$).

For strong-gravity planets with intense EUV radiation ($\Rp <\RJ$,  $\Mp>\MJ$, and $F_{\rm EUV} > 10^5\, \mathrm{\,erg/s/cm^2}$), our model underestimates the efficiency (for instance, see deviations between the purple markers and lines in Figure~\ref{fig:massloss_m} and \ref{fig:massloss_r}). This is because our model assumes an equilibrium temperature of $10^4\,\mathrm{K}$ and neglects its variation dependent on planetary parameters. In strong-gravity planets, the gas temperature exceeds $10^4\,\mathrm{K}$ by a few tens of percent due to the low density at the sonic point and the corresponding decrease in the radiative cooling rate there. Furthermore, even small temperature changes can significantly alter the mass-loss rate, as it includes an exponential factor dependent on $R_{\rm s}$, as shown in Eq.~\eqref{eq:mdot}. The effect is particularly pronounced in planets with strong gravity where the ratio of $R_{\rm s}$ to $R_{\rm p}$ is large. Therefore, our model, which assumes a constant equilibrium temperature of $10^4\,\mathrm{K}$, underestimates the efficiency of mass loss in these conditions. 
The assumption of a constant equilibrium temperature is also invalid for strong-gravity planets with low EUV radiation.  In these cases, the equilibrium temperature drops below $10^4\mathrm{\, K}$, causing our model to overestimate the efficiency. Consequently, the efficiency dependence on EUV flux differs for strong-gravity planets. This underestimation is evident from the deviations between the purple markers and lines in our results. To accurately assess the efficiency of strong-gravity planets, a detailed analysis incorporating the variability of the equilibrium temperature is essential.

We find slight deviations between our efficiencies and the conventional recombination-limited efficiencies in the weak-gravity regimes (the right part of mass-loss efficiency in Figure~\ref{fig:massloss_r} and the left part in Figure~\ref{fig:massloss_m}), even under strong EUV fluxes where the recombination-limited model is often applied. This discrepancy arises from the difference between $R_{\rm EUV}$ in Eq.~\eqref{eq:mdot1} and $\Rp$ in Eq.~\eqref{eq:mdot}.

\subsection{Comparison to radiation hydrodynamics simulations}
\label{sec:simulation}
To validate the assumptions and outcomes of our analytic model, we have performed our own one-dimensional (1D) hydrodynamics simulations with EUV photoionization heating and Ly$\alpha$ radiative cooling, which dominate the heating and cooling processes in the upper atmosphere of typical hot Jupiters.
We solve the following hydrodynamic equations:
\begin{align}
&\frac{\partial \rho}{\partial t}+\frac{1}{r^2}\frac{\partial}{\partial r} (r^2\rho v) =0\\
&\frac{\partial\rho v}{\partial t} + \rho v\frac{\partial v}{\partial r} = -\frac{\partial p}{\partial r} -\rho \frac{\partial \Psi}{\partial r}\\
&\frac{\partial E}{\partial t} + \frac{\partial Hv}{\partial r} = -\rho v \frac{\partial \Psi}{\partial r} + \rho (\Gamma-\Lambda)\\
\end{align}
where $\rho, v, p, E, H$ are gas density, velocity, pressure, energy, and enthalpy per unit volume of gas. The effective gravitational potential $\Psi$ can be given by $\Psi=-G\Mp/r-GM_*/r_*-GM_*r_*^2/2a^3$ where $a,r_*$ represents the semi-major axis and local distance to the host star as in \cite{Mitani_2022}.
The simulations include photoionization heating and recombination cooling of hydrogen in $\Gamma$ and $\Lambda$ respectively. EUV radiative transfer is handled using ray tracing. We also implement the Ly$\alpha$ cooling which dominates the radiative cooling rate. We also solve for non-equilibrium chemistry:
\begin{equation}
\frac{\partial n_{\rm H} y_i}{\partial t} + \frac{\partial n_{\rm H} y_i v}{\partial r} = n_{\rm H} R_i 
\end{equation}
where $y_i=n_i/n_{\rm H}$ and $R_i$ represent the chemical abundance and the reaction rate respectively.
Our 1D hydrodynamic simulation code is based on the CIP method \citep{Yabe_1991}.
We initialized our simulations with a hydrostatic atmosphere and at the lower boundary, we impose a fixed surface temperature ($1000\mathrm{\,K}$) and density ($n=10^{14}\mathrm{\, cm^{-3}}$), while the upper boundary allows for outflow to mimic the escaping atmosphere.
The 1D simulations employ a spatial resolution of 5000 grid points, ensuring sufficient detail to resolve the ionization front and escape flow.  
We confirm that the 1D result is consistent with the open-source 1D code ATES\citep{Caldiroli_2021}. We also confirm that the mass-loss rates of the 1D simulations are about four times higher than in 2D simulations, validating the assumption in Eq.~\eqref{eq:mdot1} as discussed in previous studies \citep{Murray-Clay_2009}.

\begin{figure*}
    \centering
    \includegraphics[width=19cm]{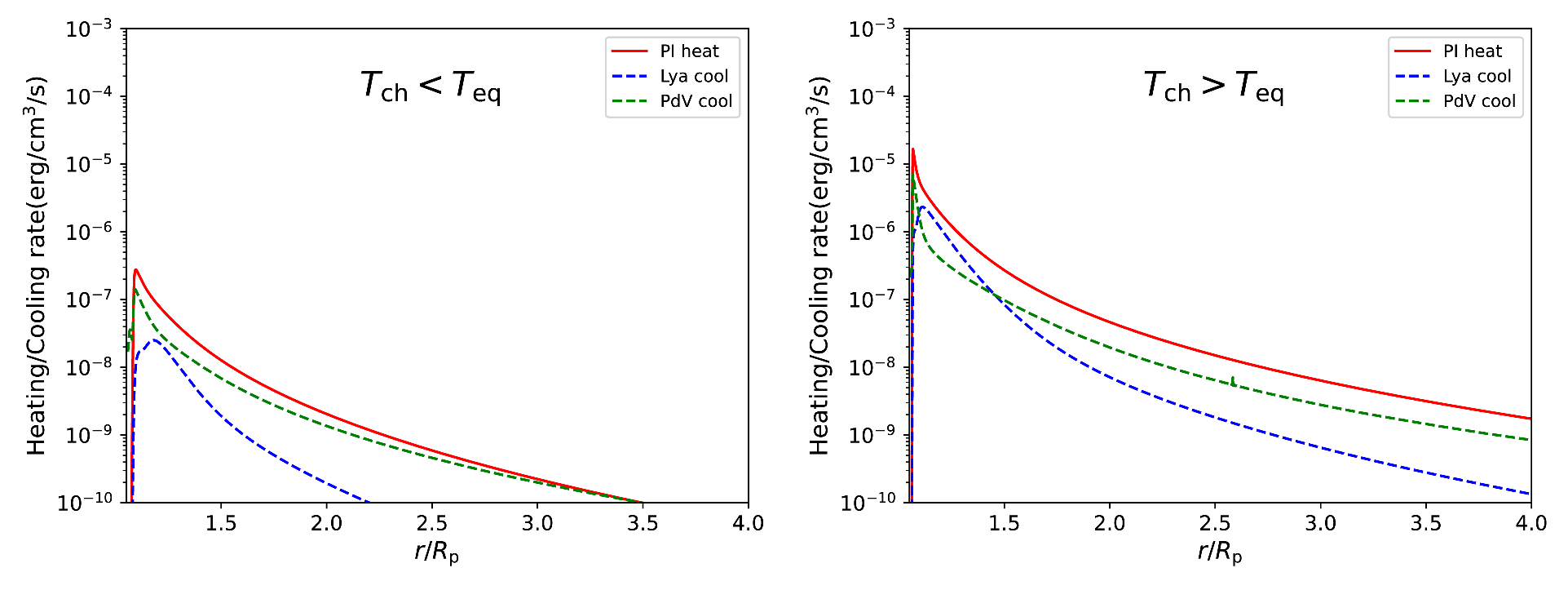}
    \caption{The heating/cooling rate profiles for $\Mp=0.7\,M_{\rm J}, \Rp=1.4\,R_{\rm J}$ with different physical conditions (left panel;$T_{\rm ch}<T_{\rm eq}$ and right panel;$T_{\rm ch}>T_{\rm eq}$). The solid curve shows the photoionization heating and the dashed curves show the radiative cooling and the PdV cooling.}
    \label{fig:heating}
\end{figure*}

Figure~\ref{fig:heating} shows the heating and cooling rate profiles for planets with $\Mp=0.7\,M_{\rm J}$ and $\Rp=1.4\,R_{\rm J}$ under different EUV fluxes ($F_0 = 1000, 50000 \mathrm{\,erg/s/cm^2}$). The higher flux is chosen so that the characteristic temperature exceeds the equilibrium temperature ($T_{\rm ch} > T_{\rm eq}$). In this case, we find that radiative cooling dominates the cooling process, and the system becomes recombination-limited. If the equilibrium temperature exceeds the characteristic temperature ($T_{\rm ch} < T_{\rm eq}$), the system becomes energy-limited. We also confirm that the gas temperature can be given by the gravitational temperature for low EUV and strong gravity planets.
This condition is consistent with previous simulations by \citet{Murray-Clay_2009}.

The simulations confirm the outcome of the analytic model that the gas temperature is consistent with the gravitational temperature if $T_{\rm ch}<T_{\rm g}$ (the lower-flux model, $F_0 = 1000\mathrm{\,erg\,cm^{-2}s^{-1}}$).
We also run two-dimensional (2D) axisymmetric hydrodynamics simulations as in our previous study \citep{Mitani_2022} and confirm that the mass-loss rates of the 1D simulations are about four times higher than in 2D simulations, validating the assumption in Eq~\eqref{eq:mdot1} as discussed in previous studies \citep{Murray-Clay_2009}.

\section{Classification and understanding of observed exoplanets}
\label{sec:classification}
Non-detections of Ly$\alpha$, H$\alpha$, and Helium triplet absorption have been found in some close-in exoplanets and the exact origin is still unknown:
Stellar wind confinement and the absence of hydrogen-dominated atmospheres can both explain the non-detection of the escaping atmosphere.
Based on the new analytic model, we can study the conditions in which non-detection can or cannot be explained by simple hydrogen-dominated atmospheric escape.

In this section, we compare our analytic model in Section~\ref{sec:Analytic} to observed close-in exoplanets. We provide the method to find peculiar exoplanets that may be affected by the stellar wind confinement, the absence of a hydrogen-dominated atmosphere, or other processes. We also provide the classification of observed exoplanets based on our essential parameters (Section~\ref{sec:Physics}) to reveal the underlying important physics in real systems.

\subsection{Observational absorption signals}
Recent observations have found some close-in planets with non-detection of hydrogen and helium absorption in their upper atmospheres. 
To demonstrate how our model can be applied to known close-in exoplanets with upper atmosphere observations, we use recent observational data from the MOPYS project \citep{Orell-Miquel_2024}. In the data, the planets are labeled as "Detection," "Non Conclusive," and "Non Detection" for both hydrogen atoms and the helium triplet.

EUV luminosity is difficult to observe in some systems. For many observed exoplanets in the data, we use XUV (1-912\AA) flux instead. For a few planets without XUV flux data, we estimate the XUV flux using the semi-major axis and the EUV luminosity of the star. We assume the age-EUV luminosity relationships from \citet{2011_Sanz-Forcada}:
\begin{equation}
    \log\, L_{\rm EUV} = 29.12 -1.24\log\,\tau
\end{equation}
where $\tau$ is the stellar age in Gyr.
We also assume that the EUV is monochromatic light with an energy of $20\,{\rm eV}$ to estimate the EUV photon flux, the cross-section of photoionization, and the deposited energy from photoionization.

Our model depends on the average of the deposited energy, the cross section, and the photon number flux. The photoionization rate depends on $F_0\sigma_0\Delta E_0$ as in Eq.~\ref{eq:gamma_euv} and the base density also depends on the EUV flux.

\begin{figure}[h]
    \centering
    \includegraphics[width=8.9cm]{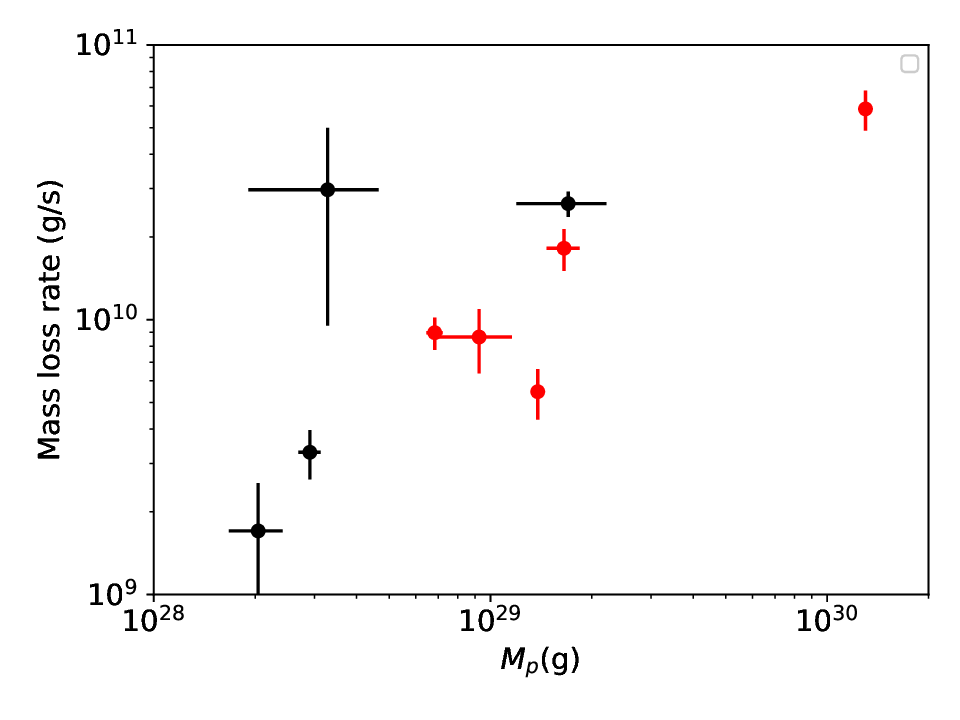}
    \caption{The distribution of close-in planets with Ly$\alpha$ observations.The red points represent the exoplanets in which hydrogen atoms have been detected in Ly$\alpha$. The black points represent the planets in which Ly$\alpha$ is not detected.}
    \label{fig:planets_Ly}
\end{figure}

\begin{figure}[h]
    \centering
    \includegraphics[width=8.9cm]{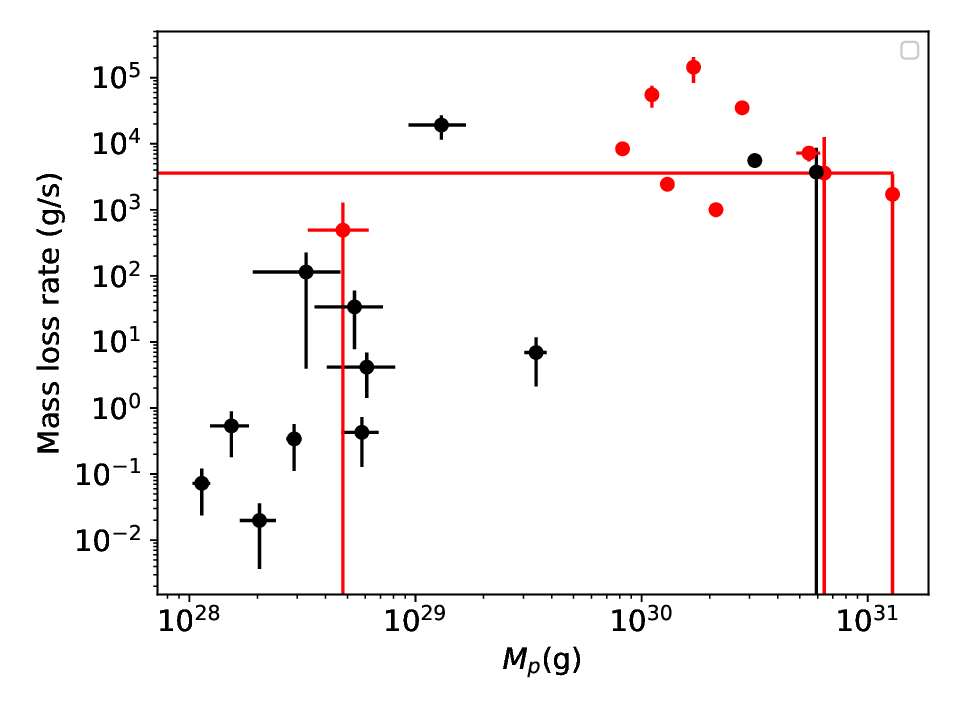}
    \caption{The distribution of close-in planets with H$\alpha$ observations. Same as Figure~\ref{fig:planets_Ly}.}
    \label{fig:planets_Ha}
\end{figure}

We estimate the mass-loss rate of neutral hydrogen $\dot{M}_{\rm HI}$ assuming the timescale of photoionization and recombination at the base determines the neutral hydrogen mass-loss rate. The neutral hydrogen mass-loss rate is one of the factors which determine the Ly$\alpha$ transit depth \citep{Owen_2023}.
Ly$\alpha$ absorption is most likely detected in intermediate EUV environments, where vigorous winds are generated while retaining sufficient atomic hydrogen against photoionization. Note that the transit depth is linearly dependent on the gas sound speed and ionization rate and only logarithmically related to the overall mass loss rate (see \cite{Owen_2023}). The photoionization rate, sound speed, and the mass-loss rate of neutral hydrogen are not independent parameters. The Lya transit depth in Owen(2023) can be a function of one of these three parameters.
If we take into account the linear dependence of the photoionization rate, the trend in the prediction of observability is similar to the neutral mass-loss rate plot because the neutral hydrogen mass-loss rate tends to be lower in the case of a high photoionization rate, and the parameter associated with the photoionization rate is also smaller. The gas velocity also does not change the trend because the difference in the sound speed of the gas is usually within a factor of two. Figure~\ref{fig:planets_Ly} shows the distribution of close-in planets with Ly$\alpha$ observations on the map of $M_\mathrm{p}$ versus the estimated mass-loss rate of the neutral hydrogen. We apply the neutral hydrogen fraction at the base. Planets with detection have indeed large mass-loss rates of neutral hydrogen. The typical error in the mass-loss rates for planets (e.g., planets with $\dot{M}_{\rm HI}<2\times10^{10}\mathrm{\,g/s}$ in Figure~\ref{fig:planets_Ly}) exposed to low EUV levels is a factor of 2. For planets with high EUV levels (planets with high mass-loss rate in Figure~\ref{fig:planets_Ha}), the error is a few tens of percent. However, the errors for highly irradiated planets might be somewhat underestimated, as we have adopted an equilibrium temperature of $10^4\Kelvin$ uniformly for these planets, neglecting variations that could arise from, e.g., atmospheric metallicity. We discuss these effects in section~\ref{sec:metal}. The error in the equilibrium temperature can be a few tens of percent, which leads to an underestimation of the mass-loss rate error by a factor of 2.

The ISM absorption prevents the observation of the Ly$\alpha$ line center and the radial acceleration due to the stellar wind or radiation pressure is necessary for the observed Lyman-alpha absorption in high-speed wings \citep{McCann_2019,Khodachenko_2019,Schreyer_2024}. Since strong stellar winds are required to accelerate significant planetary outflows, our results suggest that planets with detected Ly$\alpha$ emissions have relatively strong stellar winds and high mass-loss rates. In contrast, planets with non-detections are expected to have stellar winds that are an order of magnitude weaker compared to those with detections, or they may experience extremely strong stellar winds that confine the planetary outflow \citep{Mitani_2022}. Further modeling of the interaction between planetary outflows and stellar winds is necessary for understanding the non-detection of Ly$\alpha$ with a high mass-loss rate.

Through this analysis, we identify one outlier, K2-25 b, among the Ly$\alpha$ non-detection planets. This planet is located around the evaporation valley and may have lost its hydrogen-dominated atmosphere \citep{Rockliffe_2021}.
As demonstrated here, our model has the potential to identify peculiar cases of non-detection/detection of the upper atmosphere's absorption.

We also calculate the mass-loss rate of neutral hydrogen at the level $n=2$ using the analytical formulae of level population \cite{Christie_2013}. The population of the 2p state is set by radiative excitation and de-excitation. The population of the 2s state is set by collisional excitation from the ground state and collisional de-excitation to 2p state. 
The ratio of the population densities is given by 
\begin{equation*}
    \begin{split}
        \frac{n_{\rm 2p}}{n_{\rm 1s}} &= \frac{B_{\rm 1s\rightarrow 2p}J_{\rm Ly\alpha}}{A_{\rm 2p \rightarrow 1s}}\\
        \frac{n_{\rm 2s}}{n_{\rm 1s}} &= \frac{C_{\rm 1s\rightarrow 2s}}{C_{\rm 2s\rightarrow 2p}}
    \end{split}
\end{equation*}
where $A_{\rm 2p \rightarrow 1s}$ and $B_{\rm 1s\rightarrow 2p}$ represent the Einstein coefficients for radiative transitions, $J_{\rm Ly\alpha}$ is the radiative transition rate, and $C_{\rm i\rightarrow j}$ denotes collisional transition rates from state $i$ to state $j$. We calculate the population at the base to estimate the amount of launched excited hydrogen. It is important to note that the hydrogen population is inherently radially dependent throughout the escaping atmosphere. However, the absorption features observed in H$\alpha$ transits are primarily dominated by hydrogen located near the planetary surface \citep{Christie_2013,Mitani_2022}. Consequently, we focus on the population at the base, under the assumption that the contribution from higher altitudes has a negligible impact on the overall absorption signal.

We assume the stellar Ly$\alpha$ flux is proportional to the EUV flux. Figure~\ref{fig:planets_Ha} shows the distribution of planets with  H$\alpha$ observations. We find the clear separation around $\Mp\sim10^{30}\,\mathrm{g}$  which is consistent with the evaporation valley. 
Unlike Ly$\alpha$ observations, H$\alpha$ absorption can be more significant in an intense EUV environment because Ly$\alpha$ radiation from the host star can excite neutral hydrogen, even though the neutral hydrogen fraction becomes lower in such environments. Among the Ly$\alpha$ non-detection planets, we find one outlier, WASP-77 b. This planet is a typical hot Jupiter with a metal-poor atmosphere \citep{Smith_2024}, suggesting a high mass-loss rate due to reduced cooling (Section~\ref{sec:metal}). Despite this, no evidence of Ly$\alpha$ absorption has been detected, which may indicate that the planet is significantly affected by stellar wind confinement.    

H$\alpha$-detected planets have a large mass-loss rate of neutral hydrogen at the level $n=2$. The population of $n=2$ hydrogen depends on the stellar Ly$\alpha$ flux, which makes planets with intense EUV radiation easier to detect via H$\alpha$ absorption, unlike Ly$\alpha$ absorption. According to our model, the non-detection of neutral hydrogen can be explained by a low mass-loss rate and is not necessarily due to stellar wind confinement or the absence of a hydrogen-dominated atmosphere in most cases.
Our model can also be used to identify optimal targets for neutral hydrogen observations.

In summary, Ly$\alpha$ absorption is more likely to be detected in an intermediate EUV environment because low EUV radiation cannot drive a strong mass-loss outflow, while high EUV radiation photoionizes neutral hydrogen. H$\alpha$ absorption, on the other hand, can be detected in a high EUV environment because strong Ly$\alpha$ emission excites neutral hydrogen.

\subsection{Classification of planetary outflow}
To understand further physical conditions of observed exoplanets, we investigate relevant timescales and temperatures in observed exoplanets.

We can define the ratio between the photoionization and recombination timescales from Eqs.~\eqref{eq:t_ion} and \eqref{eq:t_rec}: 
\begin{equation}
    \frac{t_{\rm ion}}{t_{\rm rec}} = \frac{F_0\sigma}{\sqrt{\alpha_{\rm rec}/H}}.
\end{equation}

The ratio equals unity when
\begin{equation}
    \frac{F_0}{F_{\rm cr}} = \frac{\alpha_{\rm rec}\Delta E}{mc_pc_{\rm eq}c_{\rm ch}^2\sigma} \xi^{-2}.
\end{equation}

\begin{figure}[h]
    \centering
    \includegraphics[width=8.9cm]{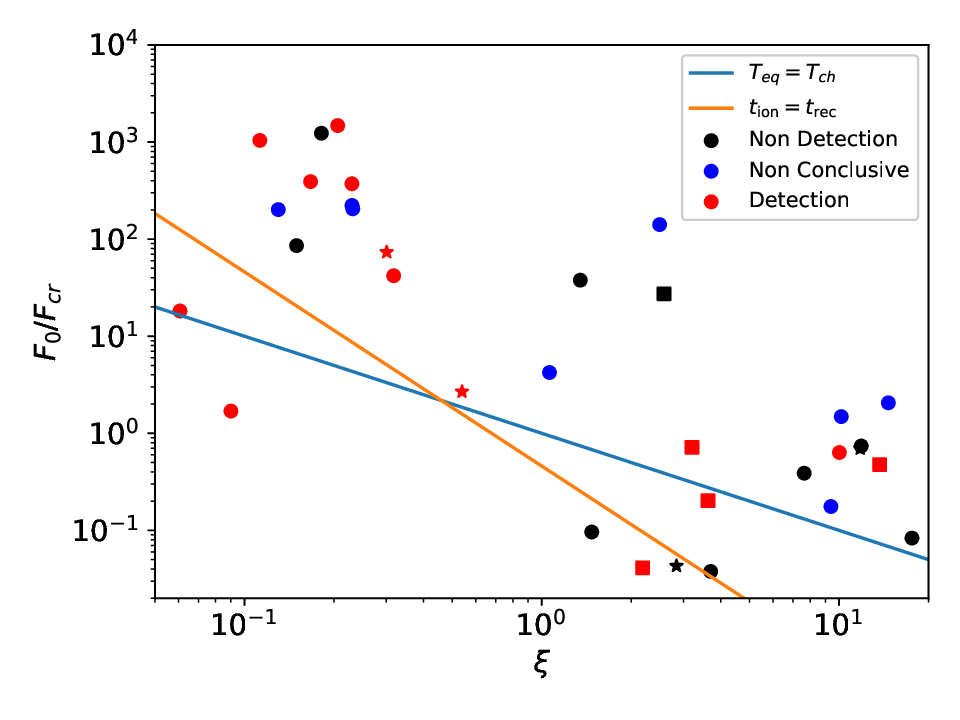}
    \caption{The distribution of close-in exoplanets with hydrogen observations. The red points represent the exoplanets in which hydrogen has been detected in Ly$\alpha$ or H$\alpha$. The blue points reflect the planets with non-conclusive detection in hydrogen absorption. The black points represent the planets in which hydrogen is not detected. Square, circle, star points represent planets with Ly$\alpha$, H$\alpha$, and both line observations respectively.  The straight lines show the condition $T_{\rm eq}=T_{\rm ch}$, $t_{\rm ion} = t_{\rm rec}$. }
    \label{fig:planets_H}
\end{figure}

The condition $t_{\rm ion} = t_{\rm rec}$ is represented by a straight line on the $\log (F_0/F_{\rm cr})$--$\log\xi$ plane in Figure~\ref{fig:planets_H}.
Similarly, $T_{\rm eq} = T_{\rm ch}$ draws another straight line. These lines divide the plane into at least three regions. Atmospheric escape is then classified into these regimes. 

Figure~\ref{fig:planets_H} shows the distribution of planets with hydrogen observations on the $\log (F_0/F_{\rm cr})$--$\log\xi$ plane. 
Above the $t_{\rm ion} = t_{\rm rec}$ line, photoionization reduces the neutral hydrogen not only in the tail but also in the launching atmosphere, making hydrogen observations difficult. This could explain why many planets have non-conclusive detections, even though the flux from their stars seems high enough to initiate photoevaporation. It may be relatively easier to detect hydrogen in planets located below the $t_{\rm ion} = t_{\rm rec}$ line, where the EUV flux exceeds the critical value. Such planets should be prioritized for future observations.
The $T_{\rm ch} = T_{\rm eq}$ line determines whether the radiative cooling dominates or not. In many close-in exoplanets with hydrogen observations, the intense radiation can heat the gas up to the equilibrium temperature.  

\begin{figure}[h]
    \centering
    \includegraphics[width=8.9cm]{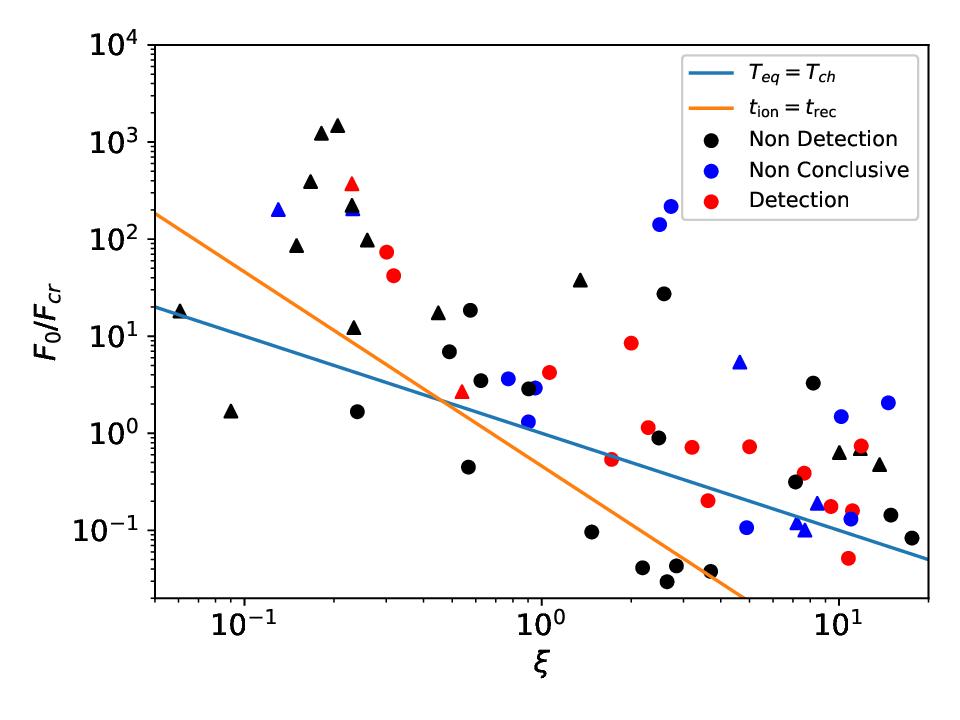}
    \caption{The distribution of close-in exoplanets with helium observations. Same as Figure \ref{fig:planets_H}. Circle/triangles represent planets with cool ($<5500\mathrm{\,K}$)/hot ($>5500\mathrm{\,K}$) host stars respectively. }
    \label{fig:planets_He}
\end{figure}

Figure~\ref{fig:planets_He} shows the distribution of planets with helium absorption. While we do not observe a clear trend, it appears that planets with helium absorption are generally in high UV  ($T_{\rm ch} > T_{\rm eq}$) environments. This correlation is expected, as high-UV conditions can lead to increased mass-loss rates and a higher fraction of helium in the triplet state. Helium absorption can be reduced by far-ultraviolet (FUV; $<13.6 \mathrm{\, eV}$) radiation due to the ionization of excited $2^3\mathrm{S}$ helium state \citep{Oklopcic_2018,Oklopic_2019}, which is not considered in this work. To consider the dependence on FUV flux, we categorize the planets based on their host star temperatures: hot ($>5500 \mathrm{\,K}$) and cool ($<5500 \mathrm{\,K}$). We find that most planets with non-detection of the helium absorption and experiencing high EUV flux have hot host stars, which likely possess higher FUV luminosities. In contrast, planets with detections and low EUV flux generally have cool host stars. This trend suggests that the EUV/FUV ratio can be a fundamental parameter of helium absorption. 
 
Further analytical investigation is required to better understand recent helium triplet observations.

It appears that classifying close-in planets based purely on the dimensionless parameters $\xi$ and $F_0/F_\mathrm{cr}$ does not reveal a trend in the detectability of hydrogen and helium escape. Instead, we need to look at the mass-loss rates of the observed species, considering their populations, as demonstrated in the previous section.
 
\section{Limitation of the model}
\label{sec:discussion}

\subsection{Evolution of mass loss regimes and planetary mass}
The mass-loss timescale for sub-Neptunian planets plays a critical role in understanding the observational features of the sub-Neptunian valley. In the early stages ($<1\mathrm{Gyr}$) of these planets, our model indicates a relatively low mass-loss efficiency ($\eta<0.1$) due to significant cooling under strong radiation ($F_{\rm EUV}>10^5\mathrm{\,erg/s/cm^3}$). As the systems age to reduce the radiation strength, the efficiency approaches unity, leading to mass-loss rates that can exceed those predicted by traditional energy-limited assumptions. Note that the planetary radius decreases with time, and the mass-loss efficiency may also be affected. The planetary radius of high-mass planets decreases by approximately 10\% over their lifetime \citep{Fortney_2010}, and the effect is not significant. For low-mass planets, the radius changes by a factor of 2 over time, but the effect can be small because the efficiency is weakly dependent on the radius in low-gravity environments, as shown in Figure~\ref{fig:massloss_m}. In total, the evolution of EUV luminosity dominates the evolution of the mass-loss rate. The mass loss during the early stage determines the total mass loss \citep[e.g.][]{Allan_2019}. Consequently, the overall mass-loss timescale is extended, aligning qualitatively with the decreasing frequency of close-in exoplanets with ages\citep{Berger_2020}.

In this study, we do not consider the impact of core-powered mass-loss. Recent theoretical models \citep{Ginzburg_2018,Owen_2024} propose that the core-powered mass-loss rate, $\dot{M}_{\rm core}$, predominates when the Bondi radius is smaller than the effective planetary radius. Under such circumstances, we can assume that the gas velocity achieves sound speeds at the effective planetary radius. If $R_{\rm s}^2\rho(R=R_{\rm s})c_{\rm inner} > R_{\rm EUV}^2 \rho_{\rm base} c_{\rm s}$, where $c_{\rm inner}$ is the sound speed in the inner atmosphere, the flow is driven by core-powered processes. Even if the EUV-driven mass-loss rate dominates the total mass loss, the core-powered flow can enhance the mass-loss flow as discussed in \cite{Owen_2024}. In cases where core-powered processes are relevant, our model would require adjustments to account for changes in the base density derived from the core-powered mass loss model. Core-powered mass loss is particularly significant in low-mass and young planets. Near the radius gap, atmospheric escape driven by EUV photoheating becomes crucial, especially as the efficiency increases in the later stages of planetary evolution.

\subsection{X-ray and metal effects}
\label{sec:metal}
In this study, we do not account for the heating effects of X-ray radiation and the cooling contributions of metal species. Recent observations have identified a hot Jupiter with a metal-rich upper atmosphere, which current classical mass-loss models fail to explain adequately. In such environments, X-rays can significantly elevate gas temperatures and penetrate deeper layers owing to their small cross-sections. For metal-poor gases, the density required to initiate atmospheric launching can be substantially higher, as it inversely correlates with the cross-section; consequently, the characteristic temperature for X-ray photoheating might be notably low. X-ray heating becomes negligible when the energy deposit of X-ray flux is smaller than that of EUV flux. Typically, the mass-loss rates driven by X-rays are lower than those driven by EUV radiation in many planetary atmospheres.

Even when incorporating the effects of metal coolants, our model is readily available once the equilibrium temperature is modified accordingly.
In cases where metals like Fe and Mg are present, metal line cooling could predominate the cooling mechanisms due to their higher abundances. For instance, the cooling rates for Mg and O are more substantial compared to other metals.

We calculate Mg II cooling which can be dominant in close-in planetary atmosphere \citep{Huang_2023}.

For real planets, many metal species contribute to the cooling rate. It is difficult to understand the physics of atmospheric escape with many metal coolants. We focus on what happens in atmosphere with one dominant metal species to investigate the physics of metal-rich atmosphere.
The Mg II cooling rate is:
\begin{equation}
    \Lambda_{\rm Mg}(\Phi_{\rm EUV}, \Mp, \Rp) = n_{\rm base} f_{\rm Mg} \frac{\Delta E}{n_l/n_u} A_{ul} \frac{n_e}{n_{\rm crit}}
\end{equation}
where $f_{\rm Mg}$ is the fraction of MgII, we use the value of excitation energy $\Delta E$, radiative decay rate $A_{ul}$, and the critical density $n_{\rm crit}$ from \cite{Huang_2017}. 
The equilibrium temperature can be calculated by solving
\begin{equation}
    \Gamma_{\rm EUV} = \Lambda_{\rm Mg}
    \label{eq:T_eq_metal}
\end{equation}
We calculate the characteristic temperature and the photoheating and cooling rates at the base using fixed planetary parameters ($M_p = 0.7 \MJ$, $R_p = 10^{10},\mathrm{\,cm}$) and varying EUV flux. We define the characteristic temperature as the metallicity-dependent equilibrium temperature when Eq.~\ref{eq:T_eq_metal} is satisfied.

\begin{figure}[h]
    \centering
    \includegraphics[width=8.9cm]{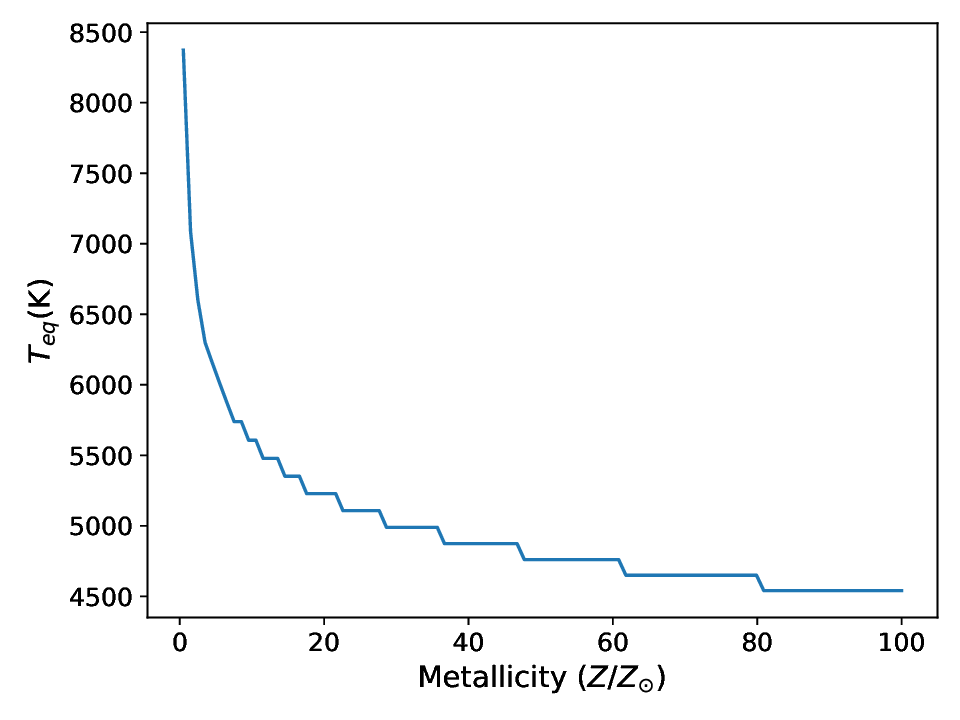}
    \caption{The metallicity dependence of the equilibrium temperature.}
    \label{fig:metal_Teq}
\end{figure}

Figure~\ref{fig:metal_Teq} shows the metallicity dependence of the equilibrium temperature. We find that the equilibrium temperature is $\sim 5000\mathrm{\,K}$ in metal-rich ($Z>10Z_{\odot}$) planets due to the strong cooling by Mg II and the mass-loss rate of such a metal-rich planet can be significantly lower than the rate of sub-solar metallicity planets.

The Ly$\alpha$ cooling rate is influenced by the electron density, as Ly$\alpha$ radiation results from hydrogen atoms being collisionally excited by electrons. Therefore, the abundance of elements like Fe, which contribute to the electron density through ionization, is also significant. However, despite these metal effects, the Ly$\alpha$ cooling rate predominantly depends on the gas temperature, maintaining the equilibrium temperature at approximately $10^4\mathrm{\,K}$ in sub-solar metallicity planets.

Observationally, the presence of metals complicates the scenario further. Metals increase the electron density through ionization, which in turn shortens the recombination timescale. This effect can be important in planets around massive stars with low EUV luminosity. The effect could enhance the observational transit depth in measurements of hydrogen neutral atoms, even though the metal cooling itself tends to reduce the hydrogen mass-loss rate.

\section{Conclusions}
\label{sec:conclusion}
EUV-driven atmospheric escape is a key process of the evolution of close-in exoplanets. In many previous planetary evolution studies, constant efficiency of mass-loss is assumed although radiation hydrodynamic simulations revealed the efficiency depends on the stellar and planetary parameters. To derive analytical formulae for the mass-loss is crucial for understanding the exoplanetary evolution.

We introduced essential parameters to describe photoheating, gas expansion, and gravitational effects within planetary atmospheres. The temperature dynamics are characterized by three distinct measures: the characteristic temperature, the equilibrium temperature, and the gravitational temperature. The characteristic temperature quantifies the intensity of photoheating and is defined as a function of the photoheating rate. The equilibrium temperature results from the balance between photoheating and radiative cooling, while the gravitational temperature reflects the influence of planetary gravity on the atmospheric structure.

Based on these quantities, we developed an analytical model for EUV-driven atmospheric escape in a hydrogen-dominated atmosphere.
Our analytical model is capable of predicting mass-loss efficiency across a broad spectrum of stellar and planetary parameters, bridging between the energy-limited (low EUV flux) and recombination-limited (high EUV flux) regimes. 

Our model indicates that the efficiency is greater than 10$\%$ for many energy-limited and low-gravity planets.
Our analysis reveals that planets exhibiting Ly$\alpha$ absorption generally possess significant mass-loss rates of neutral hydrogen, exceeding $10^{10} \mathrm{\,g/s}$. These planets typically experience an intermediate range of EUV flux—insufficient EUV flux fails to drive substantial outflows, whereas intense EUV flux leads to extensive photoionization of neutral hydrogen. Furthermore, we observe a correlation between the mass-loss rate of neutral hydrogen in the excited state ($n=2$) and the H$\alpha$ absorption signature. Planets showing H$\alpha$ absorption typically endure strong EUV flux and exhibit high mass-loss rates of excited hydrogen, often exceeding $10^3\mathrm{\,g/s}$.

Our model also identifies peculiar cases of exoplanets whose upper atmosphere observations deviate from predictions by classical EUV-driven models of hydrogen-dominated atmospheres. Such outliers may have depleted their hydrogen-dominated atmospheres or their upper atmospheres could be confined by strong stellar winds, reducing their absorption signatures.

In the future, our model can be adapted to include atmospheres with metal line cooling and other photoheating processes. These intricate physical phenomena could generalize the equilibrium/characteristic temperatures and thereby the overall mass-loss efficiency of exoplanets.

\begin{acknowledgements}
HM has been supported by International Graduate Program for Excellence in Earth-Space Science (IGPEES) of the University of Tokyo and JSPS Overseas Research Fellowship.
RK acknowledges financial support via the Heisenberg Research Grant funded by the Deutsche Forschungsgemeinschaft (DFG, German Research Foundation) under grant no.~KU 2849/9, project no.~445783058. 
RK also acknowledges financial support from the JSPS Invitational Fellowship for Research in Japan under the Fellowship ID S20156. 
We thank Naoki Yoshida for the fruitful discussions.
Numerical computations were in part carried out on the Cray XC50 at the Center for Computational Astrophysics, National Astronomical Observatory of Japan.
\end{acknowledgements}

\bibliographystyle{aa}
\bibliography{main}

\end{document}